# Electron scattering on microscopic corrugations in graphene

M. I. Katsnelson[1] and A. K. Geim[2]

[1]*Institute for Molecules and Materials, Radboud University of Nijmegen, Toernooiveld 1, 6525 ED Nijmegen, The Netherlands*

[2]*Manchester Centre for Mesoscience and Nanotechnology, University of Manchester, Oxford Road, Manchester M13 9PL, United Kingdom*

We discuss various scattering mechanisms for Dirac fermions in single-layer graphene. It is shown that scattering on a short-range potential (due to, for example, neutral impurities) is mostly irrelevant for electronic quality of graphene, which is likely to be controlled by charged impurities and ripples (microscopic corrugations of a graphene sheet). The latter are an inherent feature of graphene due to its two-dimensional nature and can also be an important factor in defining the electron mean free path. We show that certain types of ripples create a long-range scattering potential, similar to Coulomb scatterers, and result in charge-carrier mobility practically independent on carrier concentration, in agreement with experimental observations.

## Introduction

The experimental discovery of graphene, the first example of strictly two-dimensional crystals, (Novoselov et al 2004, 2005a) has led to the development of a new active research area (for review, see Geim and Novoselov 2007, Katsnelson 2007), which started expand particularly rapidly after it was demonstrated that graphene had a relativistic-like electronic spectrum with massless charge carriers (Novoselov et al 2005b, Zhang et al 2005). One of the most striking features of graphene is its very large electron mean-free path $l$, which makes graphene an enticing material for novel electronics applications (Geim and Novoselov 2007). Already in the first experiments, room-temperature ballistic transport at distances up to one micron was demonstrated (Novoselov et al 2004). Charge-carrier mobilities currently reach $\geq 15,000$ cm$^2$/Vs in single-layer graphene but further improvement of its electronic quality is desirable from the point of view of both potential applications and search for new transport phenomena (such as, for example, the fractional quantum Hall effect and other interaction phenomena that are believed to be suppressed by impurity scattering). Therefore, it is not surprising that mechanisms of electron scattering in graphene have been a subject of intensive theoretical investigation (see, e.g., Ando 2006, Nomura and MacDonald 2006, Peres et al 2006, Katsnelson 2006, Ostrovsky et al 2006, Hwang et al 2007, McCann et al 2007, and references therein). In particular, it has been shown that scattering by defects with a short-range potential having a typical radius $R \approx a$ does not significantly change graphene's resistivity $\rho$, resulting in a rather small additional contribution $\delta\rho \approx (h/4e^2)n_i R^2$ where $a$ is the interatomic distance and $n_i$ the impurity concentration (Nomura and MacDonald 2006, Ando 2006, Katsnelson and Novoselov 2007). This scattering contribution can be neglected for any feasible concentration of such defects. On the other hand, graphene's resistivity was experimentally found to be inversely proportional to concentration $n$ of charge carriers, which means that their mobility $\mu$ is almost independent on $n$ (Novoselov et al 2004, 2005b, Zhang et al 2005, Geim and Novoselov 2007). To explain this non-trivial observation, scattering on a long-range Coulomb potential due to charged impurities was invoked (Nomura and MacDonald 2006, Ando 2006, Hwang et al 2007), and it is commonly believed that Coulomb scatterers limit charge-carrier mobility in the existing graphene samples.

In this paper, we point out that in addition to charged impurities there is another source of static disorder that can lead to concentration-independent mobility $\mu$ but so far has not been taken into account. This disorder is microscopic corrugations of a graphene sheet, which are unavoidable because strictly two-dimensional (2D) crystals are extremely soft and flexible and, in fact, on the verge of structural instability (Peierls 1934, Peierls 1935, Landau 1937, Mermin 1968, Landau and Lifshitz 1980, Nelson et al 2004). The fact that graphene is actually not flat was revealed by transmission electron microscopy (TEM) where suspended graphene membranes were found to exhibit pronounced corrugations with a dominant lateral size of several nm and a height of up to one nm (Meyer et al 2007a, 2007b). Somewhat larger corrugations were earlier observed for graphene films lying on a SiO$_2$ substrate, which is attributed to the inevitable wrinkling that any thin fabric is likely to exhibit after its deposition onto a rough surface (Morozov et al 2006). Nanometre-sized ripples similar to those found in TEM were also reported in scanning-probe microscopy studies of graphene (Stolyarova et al 2007, Ishigami et al 2007). In the latter case, ripples were explained by intrinsic roughness of thermally grown SiO$_2$ (Ishigami et al 2007).

Whatever the origin of graphene's non-flatness, it must act as an additional source of scattering. In this work, we investigate this scattering mechanism and estimate its contribution to electron transport in graphene. We show that if ripples are originally created by thermal fluctuations and then become quenched (and static) during deposition on a substrate, this non-flatness should result in sufficient scattering to explain both graphene's submicron mean free path and mobility independent of concentration $n$.

**Scattering of Dirac fermions by microscopic defects with radially symmetric potential**

To appreciate specifics of scattering mechanisms for massless Dirac fermions, as compared to conventional 2D electron systems (Ando et al 1982), it is instructive to consider first the case of a small concentration of defects $n_i \ll 1/a^2$ with a radial scattering potential (Katsnelson and Novoselov 2007). Resistivity $\rho$ due to static impurities can generally be written as (Ziman 2001, Shon and Ando 1998)

$$\rho = \frac{2}{e^2 v_F^2 N(E_F)} \frac{1}{\tau(k_F)},$$
$$\frac{1}{\tau(k_F)} = n_{imp} v_F \int_0^{2\pi} d\varphi \frac{d\sigma(\varphi)}{d\varphi}(1-\cos\varphi) \quad (1)$$

where $v_F$ and $k_F$ are the Fermi velocity and wave vector, respectively, and $N(E_F) = 2k_F/\pi\hbar v_F$ is the density of states at the Fermi level (the latter expression takes into account the double-spin and double-valley degeneracy of charge carriers in graphene). The Fermi wave vector is given by $k_F = (\pi n)^{1/2}$ where $n$ is the 2D carrier concentration, $\tau$ the mean-free-path time and $\sigma(\varphi)$ the angle-dependent scattering cross section. The product $N(E_F) \cdot v_F$ is proportional to $k_F$ for both Dirac fermions and conventional 2D electrons and, therefore, any difference in transport properties of the two systems can only come though differences in their $\sigma(\varphi)$. Note that equation (1) is based on the semi-classical Boltzmann equation but its applicability for Dirac fermions has been proven by Auslender and Katsnelson (2007) who showed that, away from the Dirac point, inter-band (electron-hole) scattering processes resulted in negligible corrections to conductivity.

To determine the scattering cross section, one needs to solve the 2D Dirac equation, which for the case of massless particles and radially symmetric scattering potential $V(r)$ takes the form

$$\frac{dg_l(r)}{dr} - \frac{l}{r}g_l(r) - \frac{i}{\hbar v_F}(E-V(r))f_l(r) = 0$$
$$\frac{df_l(r)}{dr} + \frac{l+1}{r}f_l(r) - \frac{i}{\hbar v_F}(E-V(r))g_l(r) = 0 \quad (2)$$

where $l = 0, \pm 1, ...$ is the angular-momentum quantum number, and $g(r)e^{il\varphi}$ and $f(r)e^{i(l+1)\varphi}$ are components of the Dirac pseudospinor (note that the spinor index in graphene labels its two crystal sublattices rather than directions of the real spin). To be specific, we will further consider the case of a finite concentration of electrons, $E \equiv \hbar v_F k > 0$.

Modifying the standard scattering theory (Newton 1966) for our 2D case, one should try the solution of Eq.(2) outside a region of the action of the scattering potential in the form

$$g_l(r) = A[J_l(kr) + t_l H_l^{(1)}(kr)]$$
$$f_l(r) = iA[J_{l+1}(kr) + t_l H_{l+1}^{(1)}(kr)] \quad (3)$$

where the terms proportional to Bessel (Hankel) functions describe incident (scattered) waves. As a result, we obtain

$$\frac{d\sigma}{d\varphi} = \frac{2}{\pi k}\left|\sum_{l=-\infty}^{\infty} t_l e^{il\varphi}\right|^2 \quad (4)$$

The Dirac equation for massless particles (2) has an important symmetry with respect to replacements $f \leftrightarrow g$, $l \leftrightarrow -l-1$, which means $t_l = t_{-l-1}$. Accordingly, Eq.(4) can be rewritten as

$$\frac{d\sigma}{d\varphi} = \frac{8}{\pi k}\left|\sum_{l=0}^{\infty} t_l \cos[(l+1/2)\varphi]\right|^2. \quad (5)$$

Note that for $\varphi=\pi$ the cross-section is exactly zero, which means that backscattering is absent rigorously. This property was previously noticed for the case of ultrarelativistic particles in 3D (Berestetskii et al 1971). More recently, it was shown that the absence of backscattering for Dirac fermions could be responsible for their high mobility in one-dimensional channels such as carbon nanotubes (Ando et al 1998), for anomalous transport through $p$-$n$ graphene junctions (Katsnelson et al 2006, Cheianov and Falko 2006a) and, also, for Friedel oscillations around charged impurities in graphene (Cheianov and Falko 2006b).

For the case of a finite-range potential with a radius $R$ much smaller than electron wavelength $kR \ll 1$ (examples are neutral impurities and atomic-scale roughness of a substrate), all terms in Eq.(5) except for the $s$-scattering ($l = 0$) can be neglected. Then, for a random scattering potential, we generally obtain $t_0(k) \propto kR$ which yields the above mentioned estimate $\delta\rho \approx (h/4e^2)n_{imp}R^2$. So little scattering in graphene, as compared to conventional 2D systems, is a direct consequence of its Dirac-like spectrum and is largely responsible for ballistic transport even in samples unprotected from the environment (Geim and Novoselov 2007). For intuitive understanding of this effect, one may recall that light does not notice obstacles with sizes much smaller than its wavelength (Born and Wolf 1986), whereas massless Dirac electrons have exactly the same dispersion relation as photons.

Similar to light scattering, the above description breaks down in the case of resonant scattering. If the Fermi energy of Dirac fermions accidentally coincides with an energy level of an impurity, there is much stronger scattering, which in the limit of long wavelengths can be described by $t_0(k) \propto 1/\ln(kR)$ and leads to excess resistivity

$$\delta\rho \approx \frac{h}{4e^2} \frac{n_{imp}}{n \ln^2(k_F R)} \qquad (6)$$

(Katsnelson and Novoselov 2007). This means that resonant scattering could be one of the factors limiting $\mu$ in graphene (importantly, it has the same $n$-dependence as observed experimentally). In fact, electronic structure calculations (Wehling et al 2007) have shown that certain types of adsorbed molecules (e.g., $N_2O_4$) have resonant, quasi-localized levels near zero energy. It is interesting to note that Eq.(6) is also valid for the case of conventional (non-relativistic) 2D systems (Adhikari 1986). However, in the latter case, resonant scattering is not incidental, like for Dirac fermions, but the essential feature that limits electron mobility because even a weak random potential in conventional 2D systems leads to resonances near the band edge (Landau and Lifshitz 1977).

The situation is qualitatively different for charged impurities that have a long-range scattering potential. For the unscreened Coulomb potential, scattering phases for massless Dirac fermions are energy independent in the limit of low energies, which leads to resistivity inversely proportional to $n$ (Nomura and MacDonald 2006). If screening is taken into account, the scattering cross section does not change significantly except for a numerical coefficient (Nomura and MacDonald 2006, Ando 2006). Further analysis shows that for the case $\beta = \frac{Ze^2}{\hbar v_F \varepsilon} > \frac{1}{2}$ (where $Z$ is the dimensionless impurity charge and $\varepsilon$ the dielectric constant), in which the Dirac equation allows the electron to fall down at the Coulomb centre (Berestetskii et al 1971), vacuum-polarization effects diminish the initial supercritical value of $\beta$ down to a critical value of ½ (Shytov et al 2007). Nevertheless, the concentration dependence of resistivity should again remain qualitatively the same as for the unscreened Coulomb potential.

**Scattering by a generic random potential: cross-check with perturbation theory**

As shown above, in the case of a small concentration of scattering centres their contribution to resistivity can be calculated without any assumptions about the strength of the potential. In this section, we use the perturbation theory to check the applicability of the phase-scattering approach and justify it.

Let us consider a generic perturbation of the form

$$\hat{V}_{kk'} = V^{(0)}_{kk'} + \vec{\sigma}\vec{V}_{kk'} \qquad (7)$$

where $\vec{k}, \vec{k}'$ are the electron wave vectors and the Pauli matrices act on the pseudospin (sublattice) indices. This equation takes into account both scalar (electrostatic) and vector (pseudomagnetic) potentials created by defects (see, also, McCann et al 2007). Assuming that potential $V$ is small in comparison with the bandwidth and repeating the standard derivation of the Boltzmann equation in the Born approximation (Shon and Ando 1998) we find

$$\frac{1}{\tau} = \frac{4\pi}{N(E_F)} \sum_{kk'} \delta(E_k - E_F)\delta(E_{k'} - E_F)(\cos\phi_k - \cos\phi_{k'})^2 |W_{kk'}|^2 \qquad (8)$$

where $\phi_k$ is the polar angle of the wave vector $k$, $E_k = \hbar v_F k$ is the electron energy and

$$W_{kk'} = V^{(0)}_{kk'} \frac{1 + \exp[i(\phi_k - \phi_{k'})]}{2} + \frac{1}{2}\left[\left(V^{(x)}_{kk'} - iV^{(y)}_{kk'}\right)\exp(i\phi_k) + \left(V^{(x)}_{kk'} + iV^{(y)}_{kk'}\right)\exp(-i\phi_{k'})\right] \qquad (9)$$

Equation (8) corresponds to the solution of the Boltzmann equation by a variational principle (Ziman 2001) or, equivalently, by using the Mori formula for resistivity (Mori 1965), in which only intra-band matrix elements of the current operator are taken into account. For the case of a small concentration of either short-range or Coulomb scatterers, one can check by direct calculations that Eq.(8) gives the same concentration dependence of resistivity as the phase-scattering approach described in the previous section.

**Scattering by ripples**

A local curvature of a graphene sheet changes interatomic distances and angles between chemical bonds and can be described by the following nonlinear term in the deformation tensor (Nelson et al 2004)

$$\bar{u}_{ij} = \frac{1}{2}\left(\frac{\partial u_i}{\partial x_j} + \frac{\partial u_j}{\partial x_i} + \frac{\partial h}{\partial x_i}\frac{\partial h}{\partial x_j}\right) \qquad (10)$$

where $u_i$ are the components of in-plane atomic displacements and $h$ is the displacements normal to a graphene sheet. This curvature modifies the hopping integrals $\gamma$ as

$$\gamma = \gamma_0 + \left(\frac{\partial\gamma}{\partial\bar{u}_{ij}}\right)_0 \bar{u}_{ij} \qquad (11)$$

The change in nearest-neighbour hopping parameters is equivalent to the appearance of a gauge field (Morozov et al 2006) described by a "vector potential"

$$V^{(x)} = \frac{1}{2}(2\gamma_1 - \gamma_2 - \gamma_3),$$
$$V^{(y)} = \frac{1}{2}(\gamma_2 - \gamma_3), \qquad (12)$$

where the indices 1, 2 and 3 label the nearest neighbours that correspond to translational vectors $(-a/\sqrt{3},0), (a/2\sqrt{3},-a/2), (a/2\sqrt{3},a/2)$, respectively. Changes in the next-nearest-neighbour hopping also lead to an electrostatic potential $V^{(0)}$ that fluctuates in a randomly rippled graphene sheet (Castro Neto and Kim 2007). However, as follows from Eqs.(8) and (9) both vector and electrostatic potentials contribute to $\rho$ in a similar manner and, for brevity, we will further discuss only the effect of vector potential (12). This potential is equivalent to a random sign-changing "magnetic field" that was previously shown to cause additional resistivity in conventional 2D electron systems (Geim et al 1994) and suppression of weak localization in graphene (Morozov et al 2006).

For a rough estimate, Eq.(8) can be rewritten in the simplified form

$$\frac{1}{\tau} \approx \frac{2\pi}{\hbar} N(E_F) \langle \vec{V}_{\vec{q}} \vec{V}_{-\vec{q}} \rangle_{q \approx k_F} \tag{13}$$

According to Eqs.(10)-(12) the vector potential is proportional to in-plane deformations and, thus, quadratic in derivatives $\frac{\partial h}{\partial x}, \frac{\partial h}{\partial y}$. This leads to the following expression

$$\langle \vec{V}_{\vec{q}} \vec{V}_{-\vec{q}} \rangle \approx \left(\frac{\hbar v_F}{a}\right)^2 \sum_{\vec{q}_1,\vec{q}_2} \langle h_{\vec{q}-\vec{q}_1} h_{\vec{q}_1} h_{-\vec{q}+\vec{q}_2} h_{-\vec{q}_2} \rangle [(\vec{q}-\vec{q}_1)\cdot\vec{q}_1] [(\vec{q}-\vec{q}_2)\cdot\vec{q}_2] \tag{14}$$

To describe scattering on ripples for the most general case, let us assume that their height-correlation function grows with increasing distance $r$ as $\langle [h(\vec{r})-h(0)]^2 \rangle \propto r^{2H}$ where the exponent $H$ characterises the fractal dimension of ripples (Ishigami et al 2007). This expression immediately gives us the scaling behaviour of the Fourier transform correlation function $\langle |h_q|^2 \rangle \propto q^{-2(H+1)}$ as well as the $q$-dependence of $\langle \vec{V}_{\vec{q}} \vec{V}_{-\vec{q}} \rangle$ that is defined by the convolution of two functions $q^{-2H}$. The latter result corresponds to the decoupling of the four-$h$ correlation function in Eq.(14) by using the Wick theorem that is rigorous if fluctuations are Gaussian but can also be used for a qualitative estimate in other cases.

As a result, for $2H < 1$, the correlation function has a finite limit at $q = 0$

$$\langle \vec{V}_{\vec{q}} \vec{V}_{-\vec{q}} \rangle_{q=0} \approx \left(\frac{\hbar v_F}{a}\right)^2 \frac{z^4}{R^2} \tag{15}$$

where $z$ and $R$ are the characteristic height and radius of ripples. This leads to excess resistivity

$$\delta\rho \approx \frac{h}{4e^2} \frac{z^4}{R^2 a^2} \tag{16}$$

For $2H > 1$, the resistivity is proportional to $n^{1-2H}$ and, for $2H = 1$, to $\ln^2(k_F a)$.

Below we argue that ripples with $2H = 2$ can naturally occur in graphene because of the way it is prepared. Indeed, single-layer samples studied in transport experiments are currently obtained by micromechanical cleavage, a process in which individual graphene sheets detach from bulk graphite and then attach to a $SiO_2$ substrate (Novoselov et al 2005a). It is sensible to assume that during the deposition process there is a transient state in which a detached sheet behaves as a free-standing membranes and, accordingly, exhibits dynamic out-of-plane fluctuations induced by the room-temperature environment (see, e.g., Fasolino et al 2007; Abedpour et al 2007). As graphene touches the substrate, van der Waals forces instantly pin the rippled configuration (at least, partially), and any further reconstruction should be suppressed because this would require local movements along the substrate that strongly binds the atomically-thin sheet (one might imagine the deposition process as a microscopic equivalent of placing a cling film on top of a table). Therefore, the corrugations originally induced by temperature are expected to become static and, also, persist to lower temperatures without notable changes.

To estimate scattering on such ripples, we note that, in the simple harmonic approximation, the average potential energy per individual bending mode $E_{\vec{q}} = \kappa q^4 \langle |h_q|^2 \rangle / 2$ is equal to $k_B T/2$ ($\kappa \approx 1$ eV is the bending stiffness of graphene), which yields

$$\langle |h_{\vec{q}}|^2 \rangle = \frac{k_B T}{\kappa q^4} \tag{17}$$

Numerical simulations using a realistic atomic interaction potential in graphene directly confirm this result for the case of bending fluctuations with a typical length scale smaller than $l^* \approx 7-10$ nm (Fasolino et al 2007). The linear $n$-dependence of conductivity is experimentally observed for doping levels such that $k_F l^* > 1$, which justifies the use of the harmonic approximation here. We then use Eq.(17) for the pair correlation function and find the ripple resistivity as

$$\rho_r \approx \frac{h}{4e^2} \frac{(k_B T/\kappa a)^2}{n} \Lambda \tag{18}$$

where the factor $\Lambda$ is of the order of unity for $k_F l^* \cong 1$ and weakly – as $\ln^2(k_F l^*)$ – depends on $n$ for $k_F l^* \gg 1$. The above equation shows that thermally-created ripples lead to charge-carrier mobility $\mu$ practically independent on $n$, in agreement with experiments. Importantly, Eq.(18) also yields $\mu$ of the same order of magnitude as observed experimentally. One can interpret $(k_B T_q/\kappa a)^2 \approx 10^{12}$ cm$^{-2}$ as an effective concentration of static defects (ripples) induced at the quench temperature $T_q$ of 300K. We emphasize that for a weak disorder, that is, in the Born approximation, the above formalism can be used to describe electron scattering by both static (quenched) and dynamic ripples, assuming that, first, they are classical scatterers and, second, their energy at relevant $q$ is smaller than the energy of scattered Dirac fermions $\hbar q v_F$ (Ziman 2001). The former condition means that $\kappa k_F^2 a^2 << k_B T_q$ and is satisfied in the existing experiments; the latter holds if $q << \hbar v_F / \kappa$ and is even less restrictive.

For completeness, we also consider small doping levels such that $k_F l^* \leq 1$, in which the harmonic approximation can no longer be used to calculate the correlation functions. The reason for this is that thermal fluctuations with small $q$ are extremely soft, which can lead to a crumpling instability, that is, the amplitude of fluctuations normal to a graphene sheet grows linearly with increasing its size. According to a general theory of flexible membranes (Nelson and Peliti 1987, Radzihovsky and Le Doussal 1992, Nelson et al 2004) an anharmonic coupling between bending and stretching modes partially suppresses the growth of such fluctuations so that

$$\left\langle \left|h_q\right|^2 \right\rangle \approx \frac{1}{q^4}\left(\frac{q}{q_0}\right)^\eta \tag{19}$$

where $q_0 \approx \sqrt{B/\kappa} \approx 1/a$ is a typical cut-off vector at interatomic distances, $B$ the 2D bulk modulus, $\eta \approx 0.8$ the bending stiffness exponent. This should result in a concentration dependence weaker than $\rho \propto 1/n$. It should be noted however that for graphene this general theory probably does not give a complete description at smallest scales, where peculiarities of the carbon-carbon bond become essential and, for example, lead to the appearance of ripples with preferential sizes of ≈5 to 7 nm (Fasolino et al 2007). Finally, we note that, in principle, strong enough ripples can also create mid-gap (resonant) states, if the pseudomagnetic flux per ripple is larger than the flux quantum (Guinea et al 2007). According to equation (6), such prominent ripples should be very efficient scatterers. This effect cannot be described by our perturbation theory and requires further investigation.

As for comparison with experiment, the situation currently remains unclear. On the one hand, according to recent scanning probe microscopy studies of graphene on a $SiO_2$ substrate (Ishigami et al 2007) the main contribution to the height fluctuation function is due to the substrate roughness, and $2H \approx 1$. This type of bending fluctuations cannot explain the experimental behaviour of graphene's resistivity that exhibits the $\rho \propto 1/n$ dependence. Importantly, ripples with $2H \approx 1$ result only in a small value of $\delta\rho$, which is a direct consequence of their relatively short-range character, as discussed in the earlier sections. On the other hand, TEM studies of suspended membranes show that a typical height of ripples grows approximately linear with their radius (Meyer et al 2007a), which is consistent with the long-range scattering exponent $2H = 2$. Furthermore, for graphene lying on a substrate, ripples of different origins can simultaneously be present. Then, it is possible that structural characteristics of a graphene sheet could be dominated by substrate effects, whereas thermally-created ripples with their stronger, long-range scattering potential would still produce the dominant contribution into resistivity. Further experiments are required to clarify scattering mechanisms in graphene.

In conclusion, we have discussed possible mechanisms that limit charge-carrier mobility in graphene. Although charged impurities are currently considered as the most plausible source of scattering, there is an alternative explanation. We show that intrinsic corrugations of a graphene sheet create a long-range scattering potential and lead to significant resistivity that could explain the existing experimental data.

**Acknowledgements**

The work is supported by Stichting Fundamenteel Onderzoek der Materie (FOM), the Netherlands and by EPSRC (UK). We are grateful to Kostya Novoselov, Antonio Castro Neto, Annalisa Fasolino, Allan MacDonald, Sankar Das Sarma, Leonid Levitov and Patrick Lee for illuminating discussions.